\documentclass[sigconf,nonacm,dvipsnames]{aamas} 
\pdfoutput=1

\usepackage{balance} 

\usepackage{amsmath}
\usepackage{graphicx}
\usepackage{hyperref}
\usepackage{caption}
\usepackage{pgfplots}
\usepackage{rotating}
\usepackage{subcaption}
\usepackage{tikz}

\pgfplotsset{compat=1.18}
\usetikzlibrary{shapes.misc, positioning, shapes, shadows, patterns}

\title{Behaviour Modelling of Social Animals via Causal Structure Discovery and Graph Neural Networks}

\author{Gaël~Gendron$^1$ \quad Yang~Chen$^1$ \quad Mitchell~Rogers$^1$ \quad Yiping~Liu$^1$ \quad Mihailo~Azhar$^2$ \quad Shahrokh~Heidari$^1$ \quad David~Arturo~Soriano~Valdez$^1$ \quad Kobe~Knowles$^1$ \quad Padriac~O'Leary$^1$ \quad Simon~Eyre$^3$ \quad Michael~Witbrock$^1$ \quad Gillian~Dobbie$^1$ \quad Jiamou~Liu$^1$ \quad Patrice~Delmas$^1$}
\affiliation{
  \institution{
    NAOInstitute, The University of Auckland$^1$ \quad Department of Ecoscience, Aarhus University$^2$ \quad Wellington Zoo$^3$
    }
  \country{}
  }

\begin{abstract}

Better understanding the natural world is a crucial task with a wide range of applications. In environments with close proximity between humans and animals, such as zoos, it is essential to better understand the causes behind animal behaviour and what interventions are responsible for changes in their behaviours. This can help to predict unusual behaviours, mitigate detrimental effects and increase the well-being of animals.
There has been work on modelling the dynamics behind swarms of birds and insects but the complex social behaviours of mammalian groups remain less explored. In this work, we propose a method to build behavioural models using causal structure discovery and graph neural networks for time series.
We apply this method to a mob of meerkats in a zoo environment and study its ability to predict future actions and model the behaviour distribution at an individual-level and at a group level.
We show that our method can match and outperform standard deep learning architectures and generate more realistic data, while using fewer parameters and providing increased interpretability.
    
\end{abstract}

\newcommand{\BibTeX}{\rm B\kern-.05em{\sc i\kern-.025em b}\kern-.08em\TeX}

\newcommand\notindependent{\not\!\perp\!\!\!\perp}

\begin{document}

\pagestyle{fancy}
\fancyhead{}

\maketitle 

\footnotetext[1]{Corresponding author. Email address: gael.gendron@auckland.ac.nz}


\section{Introduction}

\begin{figure}
  \begin{subfigure}[t]{0.3\linewidth}
    \includegraphics[width=0.96\textwidth, height=0.96\textwidth]{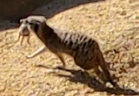}
    \caption{Carrying a pup.}
    \label{fig:behaviour_carrying}
  \end{subfigure}
  \hfill
  \begin{subfigure}[t]{0.3\linewidth}
    \includegraphics[width=0.96\textwidth, height=0.96\textwidth]{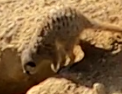}
    \caption{Digging.}
    \label{fig:behaviour_digging_burrow}
  \end{subfigure}
  \hfill
  \begin{subfigure}[t]{0.3\linewidth}
    \includegraphics[width=0.96\textwidth, height=0.96\textwidth]{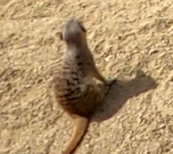}
    \caption{Low sitting or standing.}
    \label{fig:behaviour_l_sitting}
  \end{subfigure}
  \hfill
  \begin{subfigure}[t]{0.3\linewidth}
    \includegraphics[width=0.96\textwidth, height=0.96\textwidth]{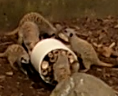}
    \caption{Interacting with a foreign object.}
    \label{fig:behaviour_object_interaction}
  \end{subfigure}
  \hfill
  \begin{subfigure}[t]{0.3\linewidth}
    \includegraphics[width=0.96\textwidth, height=0.96\textwidth]{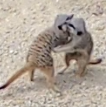}
    \caption{Playfighting.}
    \label{fig:behaviour_playfighting}
  \end{subfigure}
  \hfill
  \begin{subfigure}[t]{0.3\linewidth}
    \includegraphics[width=0.96\textwidth, height=0.96\textwidth]{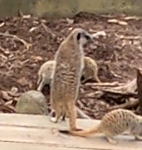}
    \caption{Raised guarding.}
    \label{fig:behaviour_raised_guarding}
  \end{subfigure}
  \hfill
  \caption{Examples of meerkat behaviours in the Meerkat Behaviour Recognition Dataset \citep{DBLP:journals/corr/abs-2306-11326}.}
  \label{fig:behaviours}
\end{figure}

Understanding and modelling non-human animal behaviour is a fundamental task in ecological research, with wide applications ranging from unusual behaviour detection \citep{buskirk1981unusual}, population dynamics \citep{morales2010building}, habitat selection analysis \citep{van2016movement}, and disease spread modelling \citep{dougherty2018going}.
One particular application of behaviour modelling is to monitor the well-being of animals in zoo environments, where evidence-based approaches for animal behaviour monitoring and prediction are essential to help zoos maximise the well-being of their captive wild animals \citep{Rose_behavioural_research, Scott_2014, Scott_group_size}.

To model animal behaviour, classical rule-based methods rely predominantly on the manual identification of correlations. In many cases, these methods begin with a hypothesis and subsequently validating it using the collected data \citep{van2016movement,buskirk1981unusual}.
However, despite the achievements of these approaches, they rely on human expertise, which can be expensive and impractical when dealing with complex behavioural rules.
The advent of machine learning has opened up the possibility of simultaneously considering multiple factors when investigating complex correlations \citep{valletta2017applications}.
However, learning correlations without recovering the cause and effect knowledge cannot provide a full understanding of the studied phenomenon \citep{DBLP:journals/pieee/ScholkopfLBKKGB21}.
While existing research has largely concentrated on the interplay between behaviour and environmental factors, the cause-and-effect relationships among behaviours have been less explored. Causal relationships among behaviours are particularly prominent in social mammals, such as meerkats and chimpanzees \citep{clutton2004behavioural,f44044be-52bf-3643-9320-44d1948f8d47,wakefield2013social} and are often interrelated in complex ways. In zoo populations, human intervention adds an additional layer of complexity. 
Consequently, determining the causal relationships between behaviours that evolve over time can be challenging.

Causality theory for time series aims to recover the causal dependencies between variables that evolve over time \citep{DBLP:journals/kais/MoraffahSKBWTRL21, DBLP:journals/corr/abs-2302-00293}. These methods are well suited to discriminate true causal links from spurious correlations (e.g. variables can be correlated because they share a common cause but have no causal links between them). A small body of work recovers the causal structure of animal behaviours but focuses on insect swarms \citep{DBLP:journals/tmbmc/Lord0OB16} or bird flocks \citep{chen2020probabilistic} and does not attempt to model the complex social interactions of individuals.

In this work, we propose an approach based on Causal Structure Discovery \citep{DBLP:journals/corr/abs-2302-00293} and Neural-Causal Inference \citep{DBLP:journals/corr/abs-2109-04173} to (1) automatically discover the causal relationships between the behaviours of individuals in a social group and (2) model and predict the behaviours of individuals over time. We build a causal graph from the data using a Causal Structure Discovery algorithm to tackle the first objective. We present the algorithm in Section \ref{sec:causal-discovery}. We then use the causal graph to constrain a Graph Neural Network \citep{DBLP:journals/tnn/ScarselliGTHM09, DBLP:conf/iclr/KipfW17} to perform Causal Inference and predict the next behaviour of an individual, as described in Section \ref{sec:causal-inference}.
In Section \ref{sec:experiments}, we apply the proposed method to simulate the behaviours of a mob of meerkats in an enclosure of the Wellington Zoo. Figure \ref{fig:behaviours} illustrates some examples of the observed behaviours.
We study the prediction abilities of the model and compare it with deep learning baselines. Then, we simulate the behaviours at both the individual and group levels, comparing the results to the ground truth.
We also discuss the advantages of the proposed method in terms of explainability. Our code is available at: \url{https://github.com/Strong-AI-Lab/behavior-causal-discovery}.
The contributions of this study are as follows:
\begin{itemize}
    \item We propose a novel interpretable method to model and predict the behaviours of social animals using causal structure discovery and graph neural networks.
    \item We apply our method to model a group of meerkats and show that it competes with standard deep learning models on behaviour prediction tasks, outperforming them in low temporal context settings.
    \item We simulate a group of meerkats over time and show that our method yields more realistic data than deep learning models.
\end{itemize}

\section{Related Work}

\paragraph{Behaviour Modelling}

To model animal behaviour, rule-based methods initially propose hypotheses and then validate them using collected data, as in \citet{van2016movement} and \citet{buskirk1981unusual}. \citet{timberlake1997animal} emphasises that these approaches can discover useful causal laws for animal behaviours but these laws can be hard to transfer to real-world situations and may carry anthropomorphic biases. \citet{sumpter2012modelling} furthermore distinguishes local and global hypotheses, i.e. hypotheses about the behaviour of a single individual against the macroscopic behaviour of a group. Approaches from neuroscience study animal behaviours from a cognitive perspective, but they cannot be applied using observational data alone \cite{SHETTLEWORTH2001277,shettleworth2009cognition}.
Automatically modelling behaviours from data with little inductive bias can mitigate these challenges.
Machine learning has emerged as a powerful tool for investigating complex rules by considering numerous factors \citep{valletta2017applications}. For instance, 
\citet{shamoun2012sensor} use decision trees to predict the behaviour of oyster-catchers from sensory data. \citet{rew2019animal} proposed an LSTM model to predict animal migration routes while considering weather and terrain.
Methods based on causality theory have also gained prominence for extracting cause-and-effect relationships from interdependent animal behaviours \citep{porfiri2018inferring,DBLP:journals/tmbmc/Lord0OB16,chen2020probabilistic}.
However, the existing work has mostly focused on easing hypothesis generation or predicting specific features, and not on building a behavioural model that could simulate data.

\paragraph{Causality}

Causality provides a theoretical framework for recovering and estimating cause and effect variables from data observations. Two popular approaches are the Potential Outcome Framework (POF) \citep{rubin1974estimating, splawa1990application, 10.1145/3444944} and the Structural Causal Model (SCM) \citep{pearl_2009, Pearl+2014+115+129, DBLP:conf/nips/XiaLBB21}. The Potential Outcome Framework usually models dependencies between only two variables. SCMs, however,  represent causal dependencies between multiple variables as a graphical model, where the variables are the nodes and the dependencies are the edges; this framework is well suited to our work. Construction of causal graphs can be divided into two tasks: \textit{Causal Structure Discovery} and \textit{Causal Inference} \citep{DBLP:journals/corr/abs-2302-00293}. The former recovers the structure of the graph, whereas the latter estimates the functions linking the effect variables to their causes.
Standard Causal Structure Discovery methods are based on statistical conditional independence tests between variables. The most noted algorithm is PC \citep{doi:10.1177/089443939100900106, spirtes2000causation}, which can recover the complete causal graph if all causal variables are observed (i.e. there are no hidden causes that can act as confounding effects). Fast Causal Inference (FCI) \citep{spirtes2000causation} and Real Fast Causal Inference (RFCI) \citep{colombo2012learning} extend PC to handle confounding effects. Time-series FCI (TsFCI) \citep{entner2010causal} adapts FCI to time-series data; variables of interest are restricted to a sliding temporal window, allowing the algorithm to see only the variables at the current and previous $\tau$ timesteps. PCMCI \citep{runge2019detecting} combines PC with Momentary Conditional Independence (MCI) tests to prune the set of variables of interest before building dependency links, allowing handling of large-scale data.
Once the causal structure is known, Causal Inference methods can be used to estimate the links between variables. Traditional approaches include do-calculus \citep{pearl2009causality}, a Bayesian inference framework, and regression methods such as Classification and Regression Trees (CART) \cite{loh2011classification}. More recent Causal Inference methods called Neural-Causal models rely on deep neural networks, in particular Graph Neural Networks \citep{DBLP:journals/corr/abs-2109-04173}, to estimate causal dependency functions \citep{DBLP:conf/nips/XiaLBB21}.

\paragraph{Graph Neural Networks}

Graph Neural Networks (GNNs) are a family of deep learning models that use a graph as an input and apply operations to update the node values \citep{DBLP:journals/tnn/ScarselliGTHM09, DBLP:conf/iclr/KipfW17}. The nodes in the graph contain feature vectors, and GNNs aggregate information from neighbouring nodes (nodes connected by an edge) to update these feature vectors. By applying multiple GNN operations, nodes can be updated based on information from distant nodes in the graph.
Popular GNN architectures include Graph Convolution Networks (GCN) \citep{DBLP:conf/iclr/KipfW17}, which apply convolution filters to graphs, Graph Attention Networks (GAT) \citep{DBLP:conf/iclr/VelickovicCCRLB18}, which use attention mechanisms to weigh the influence of neighbours on one node, and GATv2 \citep{DBLP:conf/iclr/Brody0Y22} which modifies this GAT architecture to increase the expressivity of the attention weights.

\section{Causal Behaviour Modelling}
\label{sec:model}

\begin{figure*}[ht]
    \centering
    \includegraphics[width=\linewidth]{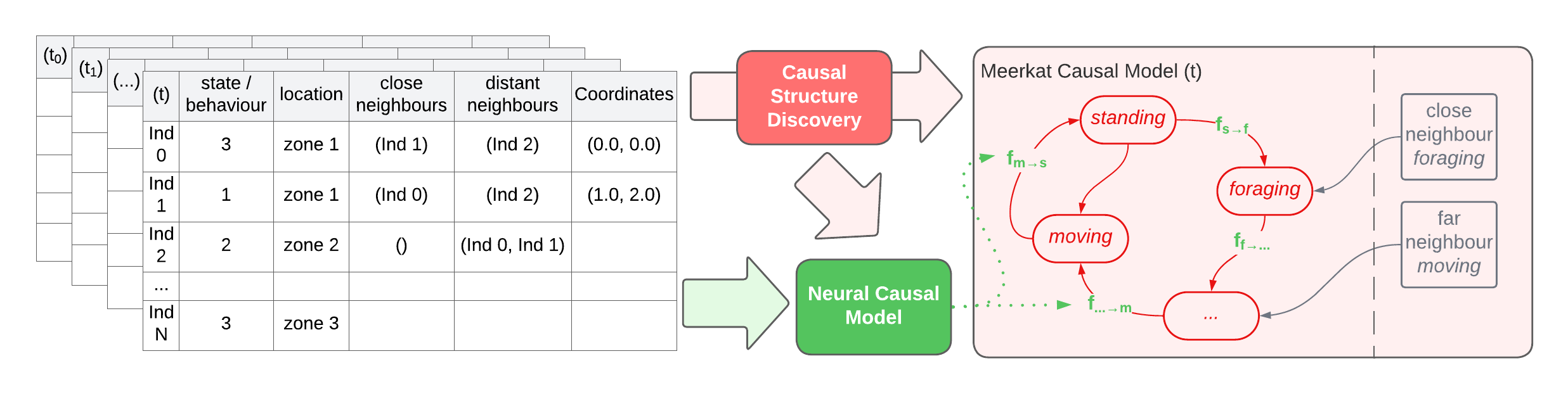}
    \caption{Causal Modelling Framework. The model is composed of two components fed from tabular data. The array contains sequences of behaviours from the studied individuals (on the left, in \textcolor{gray}{gray}). The Causal Structure Discovery module generates a causal graph that links the behaviours together (on top, in \textcolor{red}{red}). The graph contains naive transition probabilities between behaviours. These probabilities are learned by the Neural-Causal model (at the bottom, in \textcolor{ForestGreen}{green}). The network is constrained by the causal graph to follow causal paths. The resulting model combines both outputs (on the right, in \textcolor{red}{red}).}
    \Description{Causal Modelling Framework. The model is composed of two components fed from tabular data. The array contains sequences of behaviours from the studied individuals (on the left, in \textcolor{gray}{gray}). The Causal Structure Discovery module generates a causal graph that links the behaviours together (on top, in \textcolor{red}{red}). The graph contains naive transition probabilities between behaviours. The complex transition functions are learned by the Neural-Causal model (at the bottom, in \textcolor{ForestGreen}{green}). The network is constrained by the causal graph to follow causal paths. The resulting model combines both outputs (on the right, in \textcolor{red}{red}).}
    \label{fig:causal-modelling}
\end{figure*}

Modelling the behaviour of individuals from real-world data, particularly social animals, is a challenging task. Real-world data from uncontrolled environments may not contain the full scope of causes influencing individuals’ behaviour. Some uncontrolled causes may not be identifiable from available data. Moreover, the data can contain biases towards representing some behaviours more than others if the latter occur only rarely. Data scarcity also hinders the data-dependant training of deep neural networks. In particular, the data imbalance towards common behaviours makes it challenging to learn from i.i.d. data. Finally, the sequences of behaviours being highly stochastic, the absence of a known true policy complicates the evaluation of a proposed model against ground truth.

To address these challenges, we propose an approach based on Causal Structure Discovery and Causal Inference, shown in Figure~\ref{fig:causal-modelling}. We build a causal model from time-series with categorical data containing the states (i.e. behaviours) to be learned and contextual information. The causal model contains the transition functions between the states, and causal dependencies between the node variables (what nodes are causes of the transition from one state to the next). The method is divided into two modules: a \textit{Causal Structure Discovery} module that recovers the dependencies and structure of the causal graph, and a \textit{Causal Inference} module that learns the transition functions given the data and graph structure. Sections \ref{sec:causal-discovery} and \ref{sec:causal-inference} describe the architectures of the two modules.

\subsection{Causal Discovery of Behavioural Structure}
\label{sec:causal-discovery}

We aim to build a causal graph that models the links between the behaviours of social animals. We chose to represent a single individual or agent and its immediate surroundings. To simulate a multi-agent community, multiple causal models can be run concurrently. In our current work, for simplicity, we assume that all agents share the same causal processes.

The variables of interest in the graph for a timestep $t$ are the values of the possible states $S_t$ that the agent can be in (i.e. the behaviours, illustrated in Figure \ref{fig:behaviours}) and the context information $C_t$: the location of the agent, the presence of neighbours in a particular location, and the presence of neighbours doing a particular behaviour. All variables are binary and represent the presence or absence of a feature of interest. The set of variables at a given timestep is $\mathbf{X_t} = (X^1_t, \dots, X^{N+M}_t) = (S^1_t, \dots, S^N_t, C^1_t, \dots, C^M_t)$, where $N$ is the number of possible states or behaviours and $M$ is the number of context variables. We emphasise that during inference, only a subset of the learned graph is used, because only one state $S_t$ is activated and a subset of $C_t$ is activated. The causal model aims to recover the cause-effects mechanisms that lead an agent to make the decision to change (or maintain) its behaviour, i.e. the links $X^i_{t-k} \rightarrow X^j_{t+1}$. $0 < k < \tau$, where $\tau$ is the maximum size of the past time window considered, i.e. the maximum time in the past we can look back. As the context variables depend on causes not available to the system, we only aim to find a subset of links $X^i_{t-k} \rightarrow S^j_{t+1}$. The task can also be described as finding the parents $\mathcal{P}(S^j_{t+1})$ of each variable $S^j_{t+1} \in \{S^i_{t+1}\}_{i \in 1,\dots,N}$.

We use the PCMCI algorithm \citep{doi:10.1126/sciadv.aau4996} to discover the causal structure of the behaviour model. The algorithm works in two steps:

\begin{itemize}
    \item The first step consists in using the PC algorithm \citep{doi:10.1177/089443939100900106, spirtes2000causation} to prune the set of possible parent variables to a small subset of relevant parents $\hat{\mathcal{P}}(S^j_{t+1})$.
    \item The PC algorithm is not built for time-series and may contain false positives. To eliminate these, the second step uses a Momentary Conditional Independance (MCI) test to determine whether a causal link exists between the potential parent and child variable. The MCI test is defined in Equation \ref{eq:mci}.
\end{itemize}

\begin{equation}
    X^i_{t-k} \notindependent S^j_{t+1} | \hat{\mathcal{P}}(S^j_{t+1}) \setminus \{X^i_{t-k}\}, \hat{\mathcal{P}}(X^i_{t-k})
    \label{eq:mci}
\end{equation}

The MCI test verifies whether variable $S^j_{t+1}$ is independent of $X^i_{t-k}$ conditioned on the parents of $S^j_{t+1}$ (except $X^i_{t-k}$) and the parents of $X^i_{t-k}$. Observing the parents of the two variables block correlation effects. If the two variables are not independent, then  $X^i_{t-k}$ must be a cause of $S^j_{t+1}$. If $S^j_{t+1}$ or $X^i_{t-k}$ share unobserved parent variables, they can act as confounding effects and the true causal graph may not be recoverable.
An example of recovered structure is given in Section \ref{sec:explainability}.

\subsection{Behaviour Inference}
\label{sec:causal-inference}

The aim of the Causal Inference module is to estimate the function $S^j_{t+1} = f_j(\mathcal{P}(S^j_{t+1}), \epsilon^j_{t+1})$ for every variable $S^j_{t+1} \in \{S^i_{t+1}\}_{i \in 1,\dots,N}$ where $\mathcal{P}$ denotes the parents of $S^j_{t+1}$, $\mathcal{P}(S^j_{t+1}) \subset \{\mathbf{X_t}, \mathbf{X_{t-1}}, \dots, \mathbf{X_{t-\tau}}\}$, and $\epsilon^j_\mathbf{X_{t-1}}$ is random noise representing exogenous mechanisms.

\begin{figure}
    \centering
    \includegraphics[width=\linewidth]{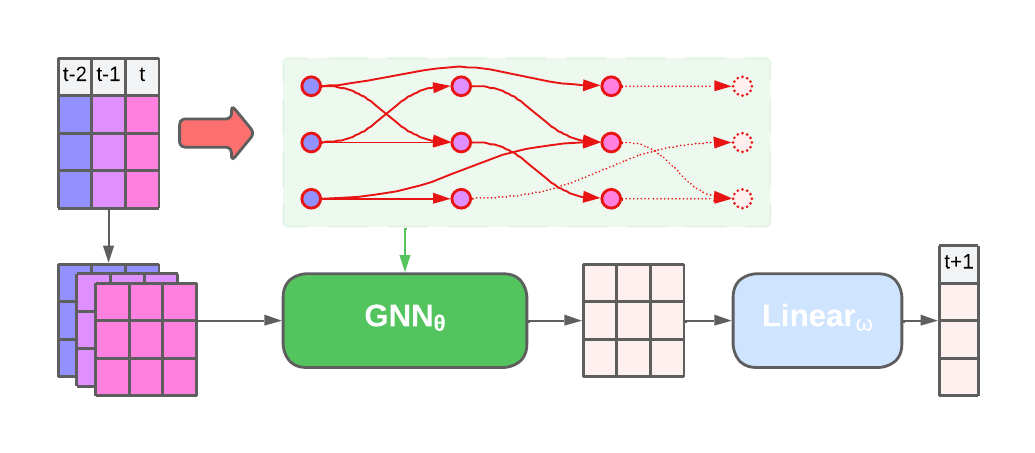}
    \caption{Neural-Causal Model based on Graph Neural Networks. The causal graph is built using the Causal Structure Discovery module (on top, in \textcolor{red}{red}) and provided to the GNN (at the bottom, in \textcolor{ForestGreen}{green}). The input values are provided as node features. The GNN aggregates the features following the causal dependencies and generates a probability vector for the next timestep using a linear layer (on the right, in \textcolor{Cyan}{blue}). }
    \label{fig:causal-inference}
\end{figure}

We use a Graph Neural Network for the Causal Inference module to take advantage of the causal structure generated by the Causal Structure Discovery module. The choice of a GNN is motivated by its ability to represent causal mechanisms under the Structural Causal Model framework \citep{DBLP:conf/nips/XiaLBB21, DBLP:journals/corr/abs-2109-04173}. Figure \ref{fig:causal-inference} shows an overview of the module architecture.

First, the input is provided to the Causal Structure Discovery module, which generates the structure of the causal model. The graph can exhibit dependencies up to $\tau$ timesteps in the past. The corresponding adjacency matrix is provided to the GNN. Input variables are formatted to be ingested by the GNN. Each binary feature of interest is converted to a one-hot encoded vector, and represents the features of a node in the graph. Because all values in the graph up to $t$ are in the data, the GNN must only compute the aggregation function for the variables at timestep $t+1$. Therefore, only a fraction of the graph is required for inference. To obtain a comprehensive overview of the model abilities, we also trained intermediate GNNs tasked with computing the intermediate values (e.g. computing $S^j_{t-\tau+1} = f_j(\mathcal{P}(S^j_{t-\tau+1}), \epsilon^j_{t-\tau+1})$). We use a final linear layer to convert the feature vectors into probabilities of variable presence. The general aggregation function of the GNN for our settings is given by Equation \ref{eq:gnn_eq}.

\begin{equation}
    S^j_{t+1} = \phi_{\mathbf{\theta_2}} \left(S^j_t, \bigoplus_{x \in \mathcal{P}(S^j_{t+1})} \gamma_{\mathbf{\theta_1}}(S^j_t, x)  \right)
    \label{eq:gnn_eq}
\end{equation}

Vector $S^j_{t+1}$ represents the features of state $j$ at timestep $t+1$, $\mathcal{P}(S^j_{t+1})$ is the set of parents of $S^j_{t+1}$ and $\bigoplus$ is a generic aggregation operator (e.g. sum $\sum$ or product $\prod$). GNN layers typically include the previous value of the variable of interest $S^j_{t}$. $\gamma$ and $\phi$ are functions parameterised by $\theta=\{\theta_1, \theta_2\}$.

\section{Application to Behaviour Prediction}
\label{sec:experiments}

In this section, we apply our method to the problem of modelling the behaviour of social animals (meerkats). We simulate individual behaviours using the proposed method. We first measure the performance of the model for direct prediction. Then, we model the behaviours of a single individual within a group, with the others evolving according to the data, and investigate how accurate the generated series is, compared to the ground truth. Then, we model the entire community and study how the group behaves as a whole. Finally, we discuss the interpretability of the model.

\subsection{Dataset and Data Preprocessing}
\label{sec:dataset}

We apply our model to the Meerkat Behaviour Recognition Dataset \citep{DBLP:journals/corr/abs-2306-11326}, a collection of 20 minutes annotated videos of a mob of meerkats in the Wellington Zoo from static cameras. The dataset tracks the position and behaviour of individual animals through video frames, and the distances between individuals is determined using an estimation of their 3D positions. Figure \ref{fig:causal-modelling} illustrates the format of the data: for every timestep, the behaviour and location of each individual is classified within pre-defined categories. 
Nearby and distant meerkats are indicated, based on the computed distances between individuals.

\begin{figure}
\centering
\begin{tikzpicture}
\begin{axis}[
    ybar,
    bar width=0.1cm,
    ymin=0,
    ylabel={Frequency},
    legend columns=2,
    symbolic x coords={
        allogroom,
        carry pup,
        dig burrow,
        foraging,
        groom,
        high sitting/standing (vigilant),
        human interaction,
        interact with pup,
        interacting with foreign object,
        low sitting/standing (stationary),
        lying/resting (stationary),
        moving,
        playfight,
        raised guarding (vigilant),
        sunbathe
    },
    xtick=data,
    xticklabel style={rotate=45,anchor=north east},
    enlarge x limits=0.1,
    width=\linewidth,
    height=5cm
]
\addplot[ybar,fill=blue!60] coordinates {
    (allogroom,0.005)
    (carry pup,0.022)
    (dig burrow,0.055)
    (foraging,0.26)
    (groom,0.009)
    (high sitting/standing (vigilant),0.084)
    (human interaction,0.0)
    (interact with pup,0.056)
    (interacting with foreign object,0.036)
    (low sitting/standing (stationary),0.124)
    (lying/resting (stationary),0.018)
    (moving,0.276)
    (playfight,0.02)
    (raised guarding (vigilant),0.029)
    (sunbathe,0.006)
};
\addlegendentry{Train data}
\addplot[ybar,fill=red!60] coordinates {
    (allogroom,0.0)
    (carry pup,0.003)
    (dig burrow,0.061)
    (foraging,0.332)
    (groom,0.005)
    (high sitting/standing (vigilant),0.082)
    (human interaction,0.014)
    (interact with pup,0.075)
    (interacting with foreign object,0.131)
    (low sitting/standing (stationary),0.062)
    (lying/resting (stationary),0.005)
    (moving,0.19)
    (playfight,0.002)
    (raised guarding (vigilant),0.03)
    (sunbathe,0.008)
};
\addlegendentry{Test data}
\end{axis}
\end{tikzpicture}
\caption{Distribution of the behaviours in the training and test data. }
\label{fig:distrib-data}
\end{figure}
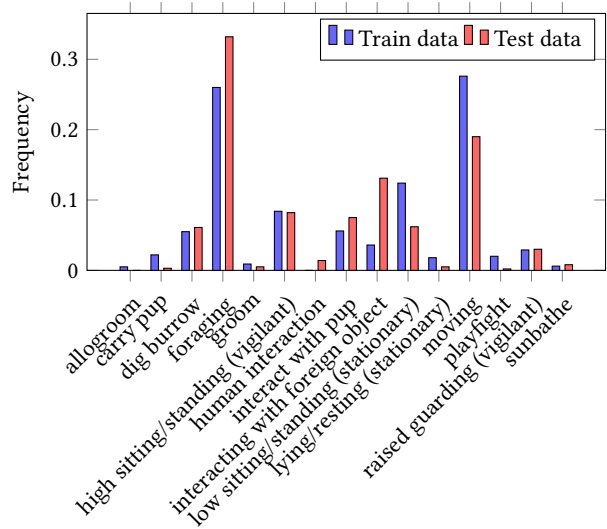

As mentioned in Section \ref{sec:model}, modelling animal behaviour using observational data is challenging. The individuals tracked in the Meerkat Behaviour Recognition Dataset express few behaviours and can maintain the same behaviour for a long period of time (for several minutes). The majority of the dataset contains sequences with little updates to the variables. The remaining sequences can be of high intensity, with behaviours changing rapidly (in less than a few seconds). Therefore, the amount of diverse data is relatively small and not independent and identically distributed (non-i.i.d.). To address this issue, sequential duplicate information was removed from the dataset. For a given timestep, we remove all the following timesteps that show no evolution in the state of the system. This removal aims to balance the dataset and remove biases towards majority behaviours. However, obtaining fully i.i.d. data is not feasible because of the very low occurrence of some of the behaviours, as shown in Figure \ref{fig:distrib-data}. We address this constraint by using our proposed causal model.

\subsection{Experimental Setup}

Our experiments aim to estimate how well our model can predict the next state or behaviour $s_{t+1}$ of an individual (i.e. the selected state among the $S^j_{t+1} \in \{S^i_{t+1}\}_{i \in 1,\dots,N}$ random variables), given its present and past states $s_t^*, s_{t-1}^*, \dots, s_{t-\tau}^*$ and the context information $C_t^*, C_{t-1}^*, \dots, C_{t-\tau}^*$. Here, $^*$ indicates the results of the ground truth policy. To study the impact of the context window size, we also ask the model to recover the intermediate values, i.e. $s_t^*$ given $s_{t-1}^*,\dots,s_{t-\tau}^*$, and $s_{t-1}^*$ given $s_{t-2}^*,\dots,s_{t-\tau}^*$, until $s_{t-\tau+1}^*$ given $s_{t-\tau}^*$. The behaviour of meerkats is stochastic, and we must use measures that take this stochasticity into account. 

We use a value of $\tau=5$. We select this value as a trade-off between the amount of information carried and the complexity of the causal discovery required. 
We perform our experiments using several GNN architectures: GCN \citep{DBLP:conf/iclr/KipfW17}, GAT \citep{DBLP:conf/iclr/VelickovicCCRLB18} and GATv2 \citep{DBLP:conf/iclr/Brody0Y22}. We use a single  GNN layer. Adding more layers would create non-causal paths in the graph. As the GNN layers aggregate information from past layers, stacking layers provides nodes with information from a greater distance in the past. This information may not be blocked by the intermediate node, which can create a non-causal backdoor path \cite{DBLP:journals/pr/ZhangSL22}.

We compare our model against two baselines: a Long Short-Term Memory (LSTM) model \citep{DBLP:journals/neco/HochreiterS97} and a Transformer model \citep{DBLP:conf/nips/VaswaniSPUJGKP17}. For a fair comparison, the LSTM has one layer and a hidden size of 128, similar to the GNNs. We use the original encoder-decoder Transformer architecture with three heads, one encoder, and one decoder layer to keep the model size as close as possible to the GNN sizes. The size of each model is listed in Table \ref{tab:param-size-models}. The GNN models have smaller sizes for inference (with $\delta\tau$ window) as we use a different GNN to compute each intermediate value. Therefore, only one of the trained GNNs is required for inference (see Section \ref{sec:causal-inference}). The LSTM and Transformer architectures perform the intermediate computations by default, so the same models are used for inference. As a result, all baseline models are larger than the proposed models.
We also build ablated models for comparison.The first model is based only on the Causal Structure Discovery module. We convert the measures of the causal strength between the variables into probabilities using softmax aggregation. The second family of ablated models contains both Causal Structure Discovery and Causal Inference modules but uses a reduced graph $\mathcal{G_{\overline{\text{corr}}}}$ instead of the full graph $\mathcal{G}$. $\mathcal{G_{\overline{\text{corr}}}}$ does not contain links in which the causal strength measures are below a statistical significance threshold. We compare the performance of the ablated models against that of the full models and baselines. All models are trained for 10 epochs with their default hyperparameters.

\begin{table}[t]
    \centering
    \caption{Number of parameters of each studied model. PCMCI is non-parametric and is omitted from the table. The first row corresponds to the total number of trained parameters, and the second row corresponds to the number of parameters used to compute next step prediction with a history of size $\tau$.  }
    \begin{tabular}{lccccc}
        \toprule
         & \textbf{GCN} & \textbf{GAT} & \textbf{GATv2} & \textbf{LSTM} & \textbf{Trans.} \\
        \midrule
        Param. & 108.9K & 109.8K & 208.8K & 113.7K & 693.7K \\
        Param. $\delta\tau$ & 26.4K & 26.6K & 43.1K & 113.7K & 693.7K \\
        \bottomrule
    \end{tabular}
    \label{tab:param-size-models}
\end{table}

\subsection{Behaviour Prediction}

In this section, we model the behaviour of a unique individual over time. We predict a sequence of behaviours $s_{t+1}, s_{t+2}, \dots$ from some initial states $s_t^*, s_{t-1}^*, \dots, s_{t-\tau}^*$ and context $ C_t^*, C_{t-1}^*, \dots, C_{t-\tau}^*$. The context remains unchanged for future predictions. E.g. $C_{t+1}^*, C_{t+2}^*$ are used to predict $s_{t+1}, s_{t+2}$. In other words, the prediction $s_{t+1}$ does not affect the context information. The other meerkats behave as expected in the ground truth.

The behaviour of meerkats is stochastic. If trained on the available data without access to the inner mechanisms of a meerkat, even a perfect policy could not reproduce the ground truth data. Therefore, we do not attempt to measure the long-term accuracy between the ground truth and the data generated by the model. We study the difference between $s_t = f(s_{t_1}, \dots, s_{t-\tau}, C_{t-1}, \dots, C_{t-\tau})$ and $s_t^*$. We report the accuracy measure on this task as a first indicator. However, as multiple behaviours can be considered valid answers for a given timestep, we also measure the Mutual Information ($I$) \citep{shannon1948mathematical,1057418} between the predicted and expected policy. The Mutual Information $I$ of two random variables corresponds to the amount of information we can obtain about one variable by observing the other, and is defined by the following equation:

\begin{equation}
    I(S_t; S_t^*) = H(S_t) - H(S_t^*|S_t)
\end{equation}

$H$ is the Shannon's entropy of a random variable. We consider the Mutual Information between the random variable representing the behaviour prediction for the current step $S_t$ and the random variable representing the true behaviour $S_t^*$. The Mutual Information returns a value between 0 and 1. Intuitively, it corresponds to the amount of uncertainty of $S_t^*$ that is removed by knowing $S_t$. 

Table \ref{tab:ind-level-results} shows the results of the experiments. The first \textbf{Acc.} column includes the computation of the intermediate values. The full-size model is used for these computations. The \textbf{Acc. $\delta\tau$} and \textbf{Mutual I.} metrics use the final output only and only require the smaller version of the model. Unsurprisingly, the PCMCI performs consistently worse than all models on every metric. For the accuracy metric on the full prediction, PCMCI+GCN$_{\mathcal{G}}$ performs the best. PCMCI+GCN$_{\mathcal{G_{\overline{\text{corr}}}}}$ achieves similar accuracy, within the standard deviation interval. Overall, the models with the ablated graph perform similarly to their counterparts with the full graph. This highlights that the causal paths contain the necessary information for the task, and removing the correlation paths does not significantly impact performance.
PCMCI+GATv2$_{\mathcal{G}}$ also performs well, achieving an accuracy similar to that of the LSTM. The Transformer model achieves a lower accuracy than all the other models, with the exception of PCMCI. As Transformers often require a large quantity of data to learn, data scarcity is a possible reason for this poor performance. Interestingly, the GNN models obtain greater accuracy on the intermediate task, with lower temporal context. 
In the final prediction task (with $\delta\tau$ context), the LSTM achieves the highest accuracy. PCMCI+GCN$_{\mathcal{G}}$ was the second-best model. The LSTM also achieves the highest Mutual Information, although the scores of all the models are very low. PCMCI+GCN$_{\mathcal{G}}$ and the Transformer achieves similar $I$ scores.

The models can approximate the true behaviour of an individual agent in more than half of the cases but the low Mutual Information scores show that the models contain little information regarding the causal mechanisms of the agents. We also observe that the proposed Neural-Causal models can yield similar or better performance than well-established models for a significantly smaller size network.

\begin{table}[t]
    \centering
    \caption{Performance of the models on the individual-level prediction tasks. The mean and standard deviation across 3 training runs is shown. \textit{Acc.} is the accuracy for the intermediate and final timesteps. \textit{Acc. $\delta\tau$} is the accuracy for the last prediction only (with a history of size $\tau$). \textit{Mutual I.} is the Mutual Information associated to the $\delta\tau$ prediction. For all metrics, the higher the better. The best results are indicated in \textbf{bold} and the second best in \textit{italics}. }
    \begin{tabular}{lccc}
        \toprule
         & \textbf{Acc.} & \textbf{Acc. $\delta\tau$} & \textbf{Mutual I.} \\
        \midrule
        PCMCI$_{\mathcal{G}}$ & 0.287 $\pm$0.000 & 0.471 $\pm$0.000 & 0.004 $\pm$0.000 \\ 
        PCMCI+GCN$_{\mathcal{G}}$ & \makebox[0pt][c]{\textbf{0.588 $\pm$0.008}} & \textit{0.517 $\pm$0.005} & 0.143 $\pm$0.025 \\ 
        PCMCI+GAT$_{\mathcal{G}}$ & 0.482 $\pm$0.015 & 0.308 $\pm$0.045 & 0.111 $\pm$0.009 \\ 
        PCMCI+GATv2$_{\mathcal{G}}$ & 0.567 $\pm$0.012 & 0.450 $\pm$0.017 & 0.116 $\pm$0.004 \\ 
        PCMCI$_{\mathcal{G_{\overline{\text{corr}}}}}$ & 0.287 $\pm$0.000 & 0.471 $\pm$0.001 & 0.004 $\pm$0.000 \\ 
        PCMCI+GCN$_{\mathcal{G_{\overline{\text{corr}}}}}$ & \textit{0.582 $\pm$0.009} & 0.511 $\pm$0.006 & 0.126 $\pm$0.000 \\ 
        PCMCI+GAT$_{\mathcal{G_{\overline{\text{corr}}}}}$ & 0.494 $\pm$0.019 & 0.300 $\pm$0.032 & 0.125 $\pm$0.008 \\
        \makebox[0pt][l]{PCMCI+GATv2$_{\mathcal{G_{\overline{\text{corr}}}}}$} & 0.548 $\pm$0.014 & 0.434 $\pm$0.009 & 0.107 $\pm$0.015 \\ 
        \hline
        LSTM & 0.565 $\pm$0.007 & \makebox[0pt][c]{\textbf{0.547 $\pm$0.011}} & \makebox[0pt][c]{\textbf{0.190 $\pm$0.009}} \\ 
        Transformer & 0.343 $\pm$0.068 & 0.332 $\pm$0.065 & \textit{0.145 $\pm$0.011} \\
        \bottomrule
    \end{tabular}
    \label{tab:ind-level-results}
\end{table}

\subsection{Individual-Level Simulation}

In this section, we aim to study the behaviour of an individual when simulated for a longer period of time. Using our models, we generated new time-series data of the same length as the available ground-truth time-series. The initial state and context information are extracted from the ground truth. The models then run partially autoregressively. Indeed, as in the previous section, we focus on the behaviour of a single individual while the rest of the group is unaffected. We generate a sequence $s_{t+1}, s_{t+2}, \dots$ while the context $C_{t+1}^*, C_{t+1}^*, \dots$ remains unchanged.

\paragraph{Clustering}

\begin{figure*}[t]
  \begin{subfigure}[t]{0.19\linewidth}
    \centering
    \includegraphics[width=0.96\textwidth,trim=3cm 0.5cm 19.5cm 6.75cm,clip]{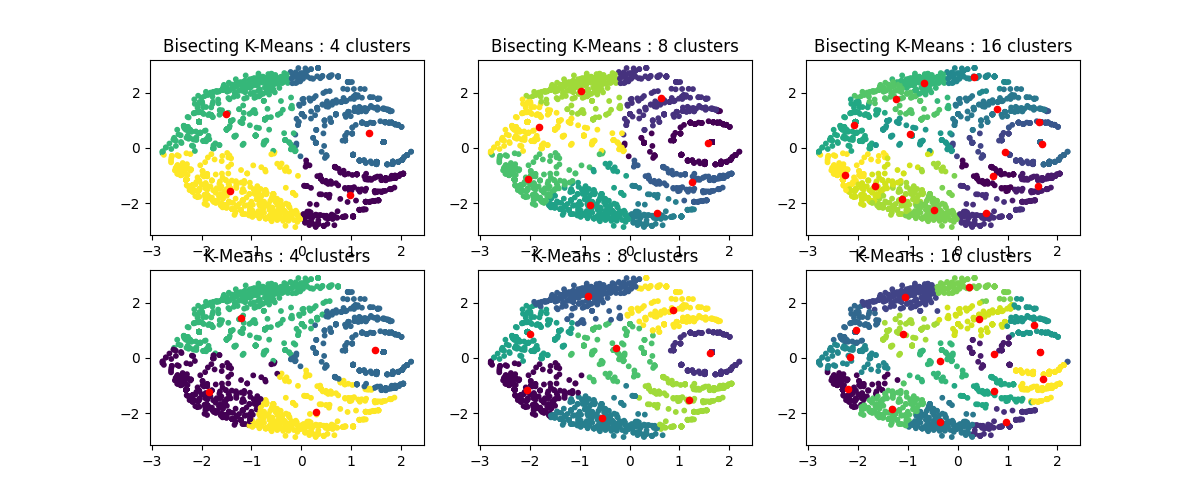}
    \caption{Ground truth.}
    \label{fig:ground_truth_cluster}
  \end{subfigure}
  \hfill
  \begin{subfigure}[t]{0.19\linewidth}
    \includegraphics[width=0.96\textwidth,trim=3cm 0.5cm 19.5cm 6.75cm,clip]{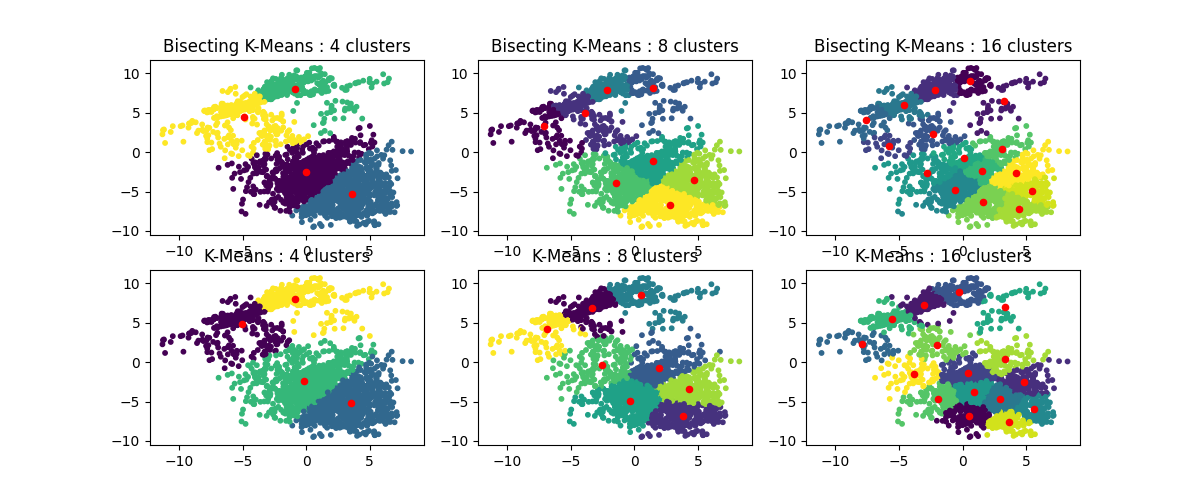}
    \caption{PCMCI.}
    \label{fig:pcmci_cluster}
  \end{subfigure}
  \hfill
  \begin{subfigure}[t]{0.19\linewidth}
    \includegraphics[width=0.96\textwidth,trim=3cm 0.5cm 19.5cm 6.75cm,clip]{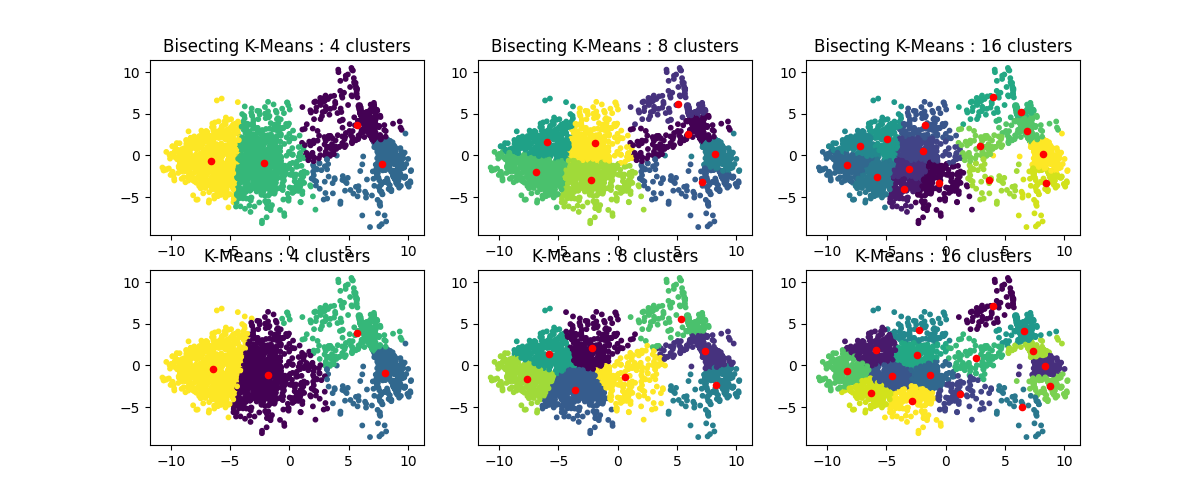}
    \caption{PCMCI+GCN.}
    \label{fig:pcmci_gcn_cluster}
  \end{subfigure}
  \hfill
  \begin{subfigure}[t]{0.19\linewidth}
    \includegraphics[width=0.96\textwidth,trim=3cm 0.5cm 19.5cm 6.75cm,clip]{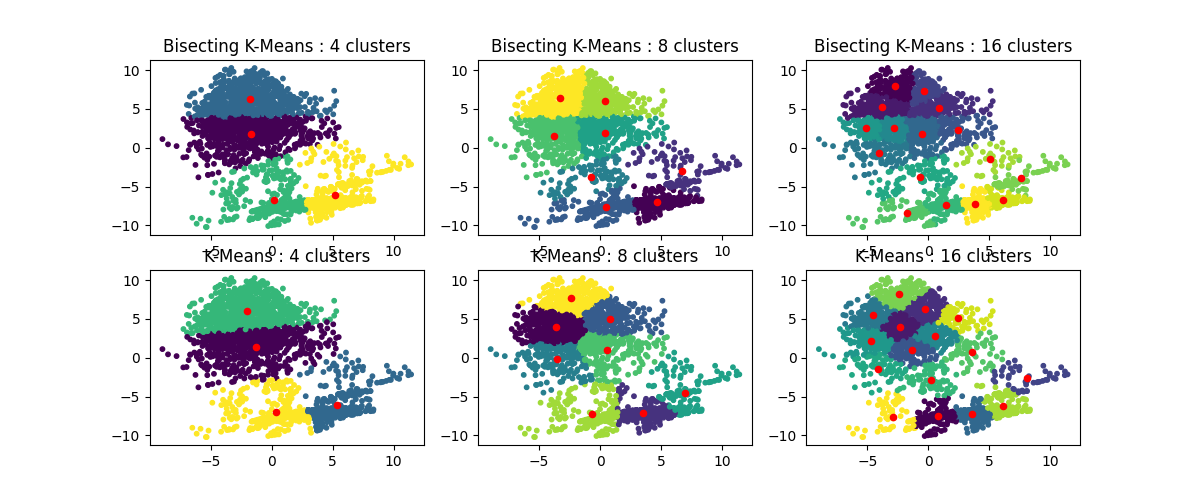}
    \caption{PCMCI+GAT.}
    \label{fig:pcmci_gat_cluster}
  \end{subfigure}
  \hfill
  \begin{subfigure}[t]{0.19\linewidth}
    \includegraphics[width=0.96\textwidth,trim=3cm 0.5cm 19.5cm 6.75cm,clip]{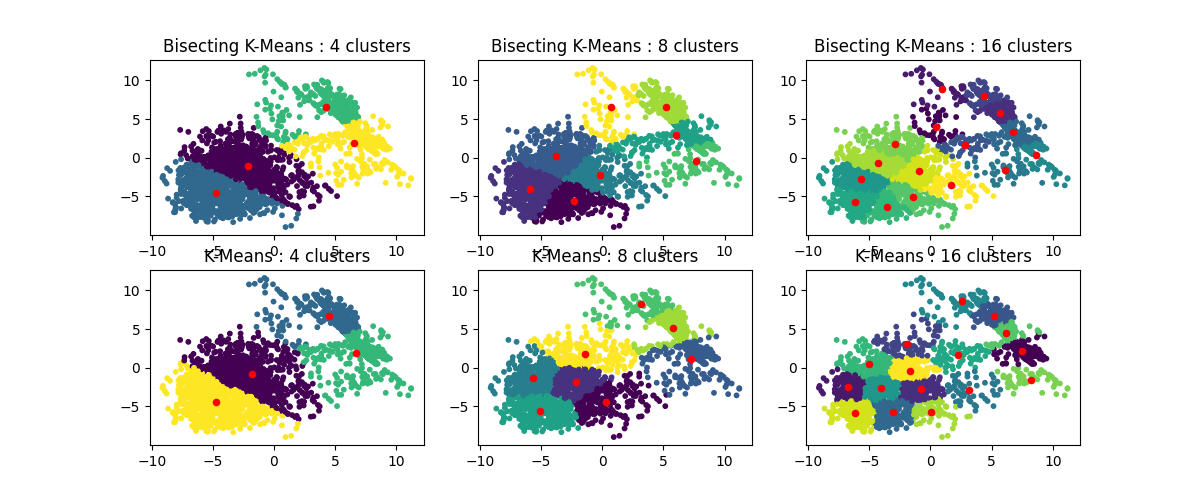}
    \caption{PCMCI+GATv2.}
    \label{fig:pcmci_gatv2_cluster}
  \end{subfigure}
  \hfill
  \begin{subfigure}[t]{0.19\linewidth}
    \includegraphics[width=0.96\textwidth,trim=3cm 0.5cm 19.5cm 6.75cm,clip]{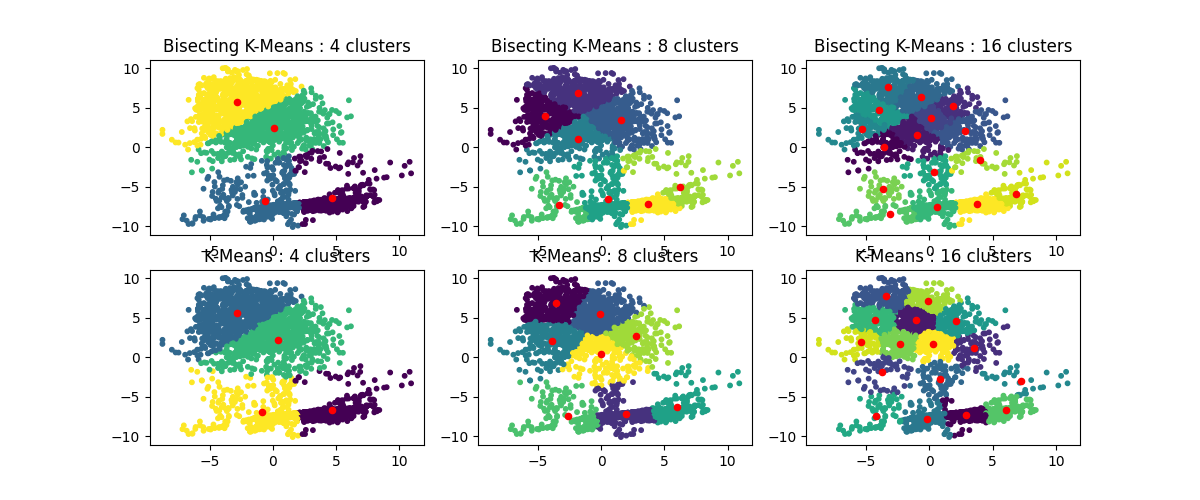}
    \caption{LSTM.}
    \label{fig:lstm_cluster}
  \end{subfigure}
  \hfill
  \begin{subfigure}[t]{0.19\linewidth}
    \includegraphics[width=0.96\textwidth,trim=3cm 0.5cm 19.5cm 6.75cm,clip]{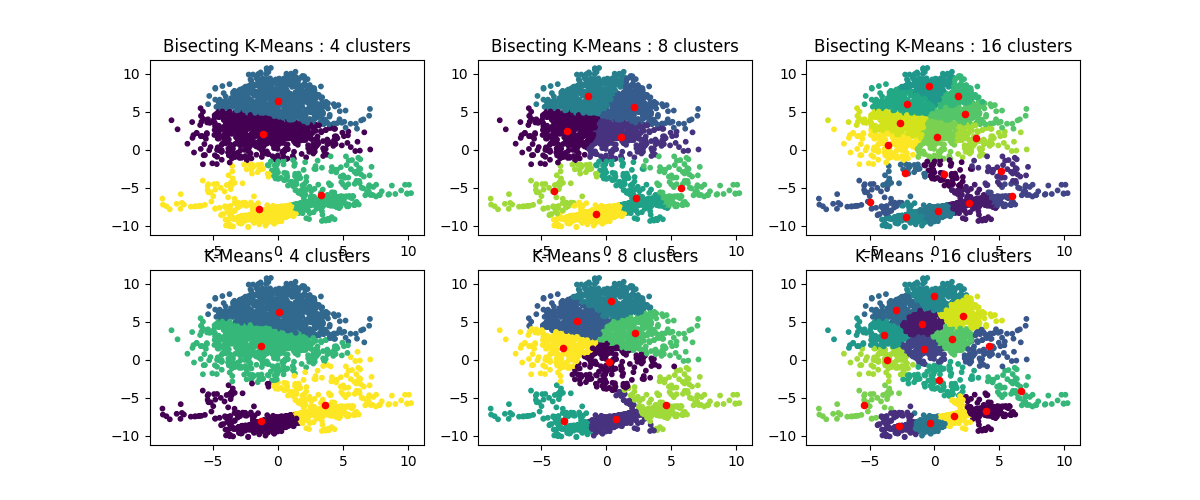}
    \caption{Transformer.}
    \label{fig:tf_cluster}
  \end{subfigure}
  \hfill
  \begin{subfigure}[t]{0.19\linewidth}
    \includegraphics[width=0.96\textwidth,trim=3cm 0.5cm 19.5cm 6.75cm,clip]{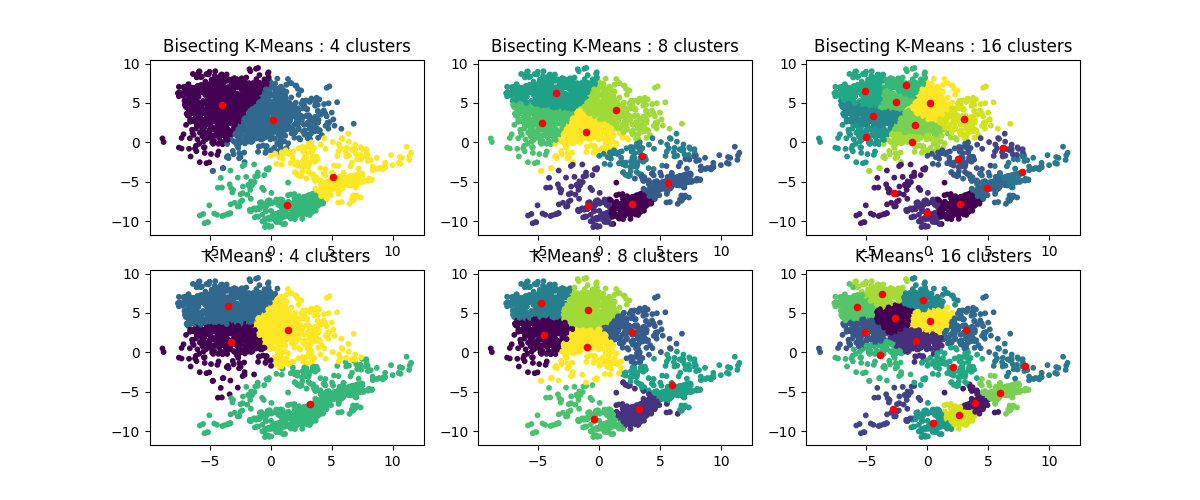}
    \caption{PCMCI$_C$.}
    \label{fig:pcmci_community_cluster}
  \end{subfigure}
  \hfill
  \begin{subfigure}[t]{0.19\linewidth}
    \includegraphics[width=0.96\textwidth,trim=3cm 0.5cm 19.5cm 6.75cm,clip]{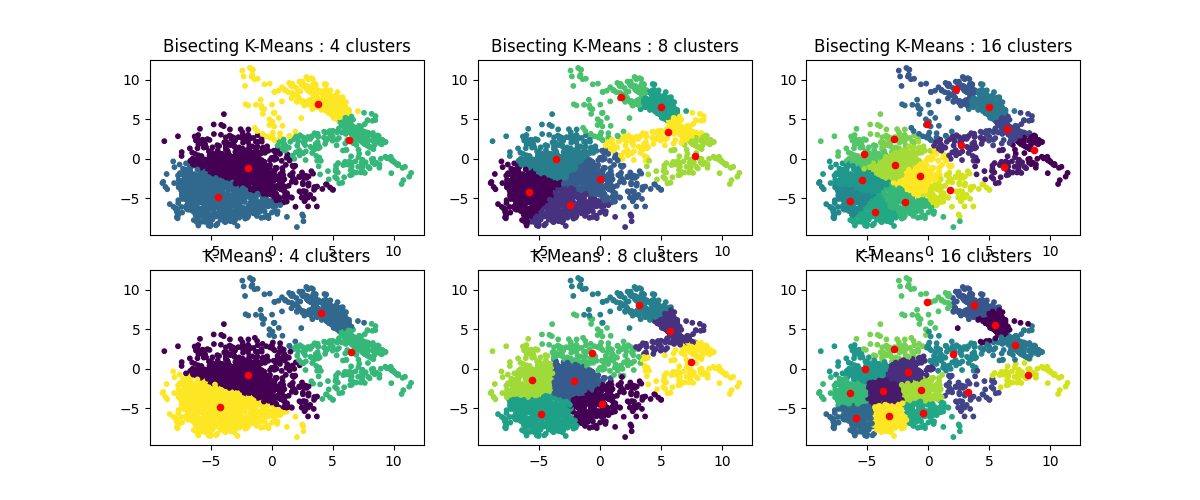}
    \caption{PCMCI+GCN$_C$.}
    \label{fig:pcmci_gcn_community_cluster}
  \end{subfigure}
  \hfill
  \begin{subfigure}[t]{0.19\linewidth}
    \includegraphics[width=0.96\textwidth,trim=3cm 0.5cm 19.5cm 6.75cm,clip]{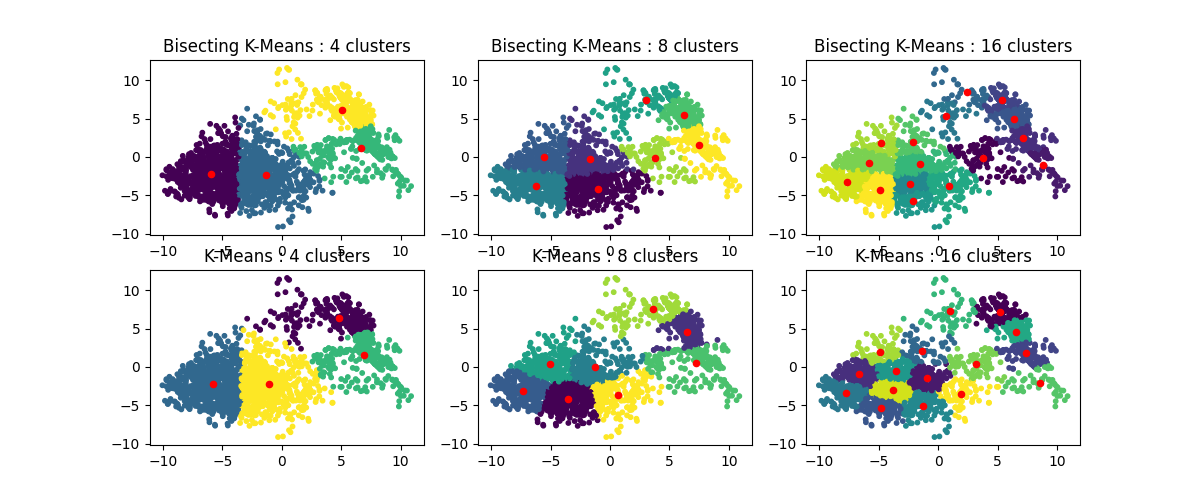}
    \caption{PCMCI+GAT$_C$.}
    \label{fig:pcmci_gat_community_cluster}
  \end{subfigure}
  \hfill
  \begin{subfigure}[t]{0.19\linewidth}
    \includegraphics[width=0.96\textwidth,trim=3cm 0.5cm 19.5cm 6.75cm,clip]{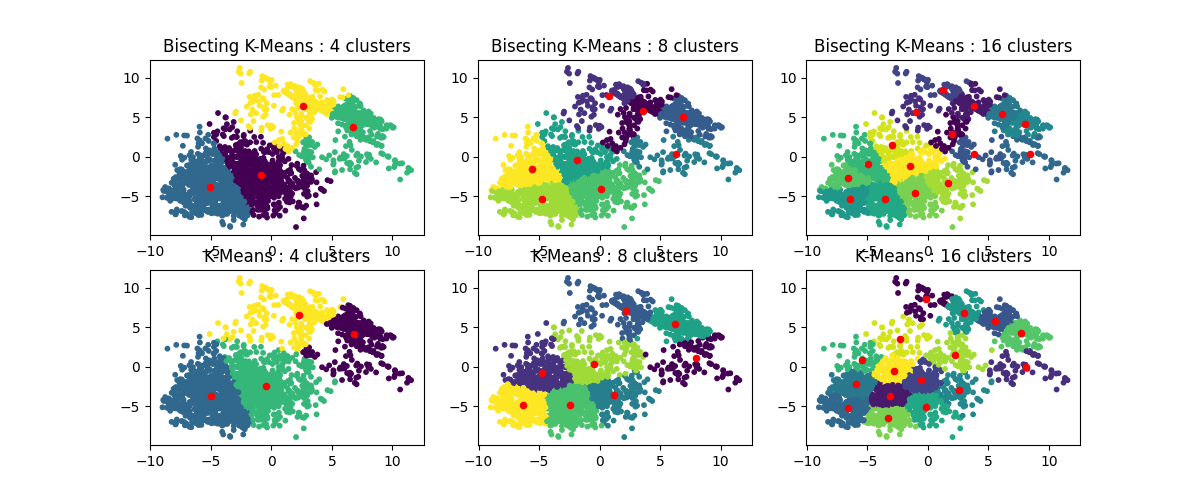}
    \caption{PCMCI+GATv2$_C$.}
    \label{fig:pcmci_gatv2_community_cluster}
  \end{subfigure}
  \begin{subfigure}[t]{0.19\linewidth}
    \includegraphics[width=0.96\textwidth,trim=3cm 0.5cm 19.5cm 6.75cm,clip]{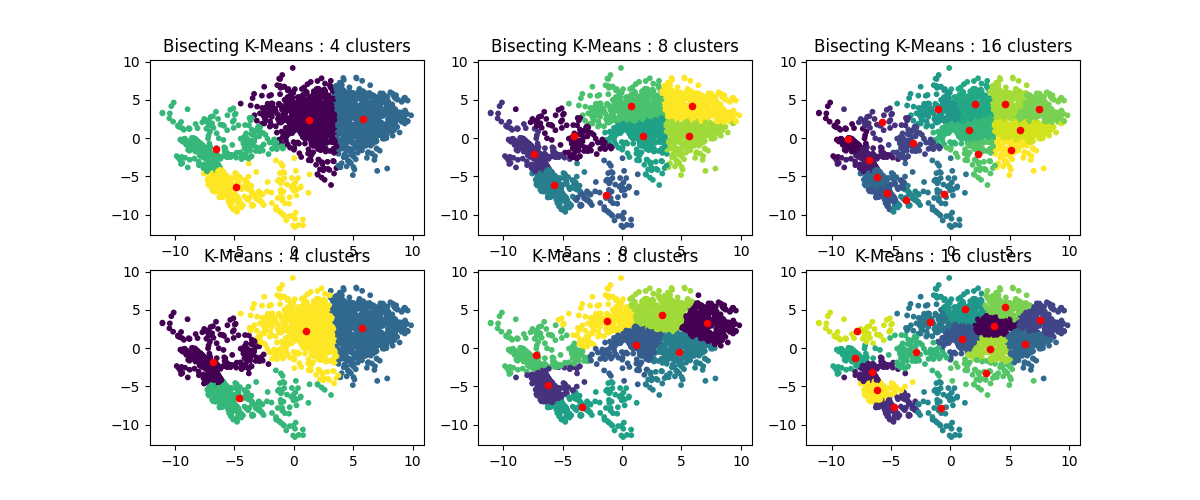}
    \caption{LSTM$_C$.}
    \label{fig:lstm_community_cluster}
  \end{subfigure}
  \begin{subfigure}[t]{0.19\linewidth}
    \includegraphics[width=0.96\textwidth,trim=3cm 0.5cm 19.5cm 6.75cm,clip]{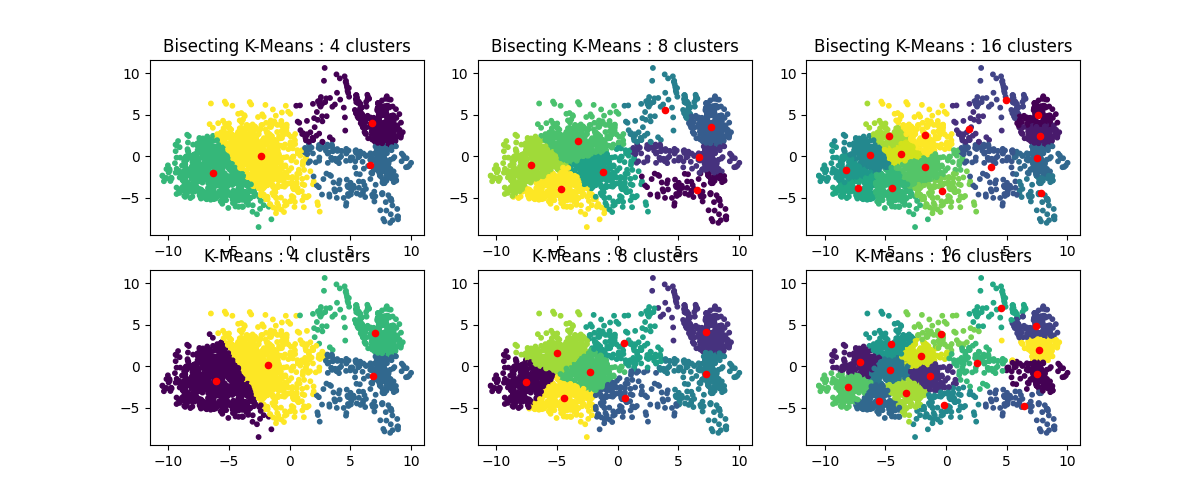}
    \caption{Transformer$_C$.}
    \label{fig:tf_community_cluster}
  \end{subfigure}
  \hfill
  \caption{Projection on a 2D space and K-Means clustering of the generated and true data samples. The ground truth is shown on the top left corner. The results of individual-level simulations are shown from Figures \ref{fig:pcmci_cluster} to \ref{fig:tf_cluster}. The results of group-level simulations are shown from Figures \ref{fig:pcmci_community_cluster} to \ref{fig:tf_community_cluster} and indicated with a $_C$. }
  \label{fig:clusters}
\end{figure*}

We sample short snippets (of $\tau+1$ timesteps) from each time-series and concatenate the vectors of each timestep to build a single vector embedding of the snippet. We project the resulting vector on a 2D space with Multidimensional Scaling (MDS) \citep{borg2005modern} and categorise the data into $K=4$ clusters with K-Means. The objective is to observe whether the generated short sequences of behaviours follow similar patterns to those in the ground truth. The results are shown in Figure \ref{fig:clusters}.
The visualisation obtained for the ground truth data are very different from that obtained with the simulations. The vast majority of the true video snippets are well separated from each other, and the samples in the cluster on the right of the image are structured as concentric circles. In contrast, the simulated samples do not form any organised structures. All models generate similar patterns. The samples from all the models are either concentrated into a large convex shape or into one or two smaller shapes. This result visually illustrates the lack of information regarding the inner mechanisms of the individual agents, as shown in the Mutual Information score of Table \ref{tab:ind-level-results}.

\paragraph{Discrimination}

To quantify the difference between the ground truth and the simulated behaviours, we trained a discriminator model to classify true and counterfeit data. Table \ref{tab:discriminator_results} shows these results. We focus on the first half of the table. On average, the LSTM discriminator reaches a higher accuracy than the Transformer discriminator. The accuracy is also more consistent, with a systematically smaller standard deviation by one order of magnitude. As for the Transformer prediction model, data scarcity is a likely explanation for the lower discrimination performance. We now focus on the performance of the generation models to fool the discriminator. The GNN models achieve better performance than the baseline prediction engines with the LSTM discriminator, with PCMCI+GATv2$_{\mathcal{G}}$ achieving the best score. The LSTM achieves the best performance with the Transformer discriminator. The GNN models perform better on the most challenging discriminator, indicating that they generated more plausible data than the baseline models. Surprisingly, PCMCI achieved a better score than the LSTM and Transformer. To generate plausible data, knowing the causal mechanisms seems to be more important than accurately predicting the next timestep.

\begin{table}[t]
    \centering
    \caption{Accuracy of the discriminator. The mean and standard deviation across 3 training runs is shown. The discriminator is tasked to distinguish real samples from simulated samples. We evaluate the ability of the models to generate samples similar to the true distribution and fool the discriminator. Therefore, the lower accuracy the better. The best results are indicated in \textbf{bold} and the second best in \textit{italics}. }
    \begin{tabular}{cccccc}
        \toprule
         & \textbf{LSTM-Discr.} & \textbf{Transformer-Discr.} \\
        \midrule
        PCMCI$_{\mathcal{G}}$ & 0.888 $\pm$0.004 & 0.892 $\pm$0.039 \\ 
        PCMCI+GCN$_{\mathcal{G}}$ & \textit{0.875 $\pm$0.001} & 0.823 $\pm$0.098 \\ 
        PCMCI+GAT$_{\mathcal{G}}$ & 0.880 $\pm$0.007 & 0.880 $\pm$0.018 \\ 
        PCMCI+GATv2$_{\mathcal{G}}$ & \textbf{0.874 $\pm$0.003} & 0.893 $\pm$0.039 \\ 
        LSTM & 0.893 $\pm$0.005 & \textbf{0.786 $\pm$0.161} \\ 
        Transformer & 0.892 $\pm$0.004 & \textit{0.818 $\pm$0.077} \\ 
        \hline
        PCMCI$_{\mathcal{G}C}$ & 0.873 $\pm$0.005 & \textit{0.591 $\pm$0.043} \\ 
        PCMCI+GCN$_{\mathcal{G}C}$ & \textit{0.866 $\pm$0.002} & \textbf{0.543 $\pm$0.060} \\ 
        PCMCI+GAT$_{\mathcal{G}C}$ & \textbf{0.865 $\pm$0.002} & 0.660 $\pm$0.021 \\ 
        PCMCI+GATv2$_{\mathcal{G}C}$ & \textit{0.866 $\pm$0.001} & 0.839 $\pm$0.041 \\ 
        LSTM$_C$ & 0.888 $\pm$0.002 & 0.649 $\pm$0.150 \\ 
        Transformer$_C$ & 0.887 $\pm$0.001 & 0.732 $\pm$0.164 \\ 
        \bottomrule
    \end{tabular}
    \label{tab:discriminator_results}
\end{table}

\begin{figure*}[t]
  \begin{subfigure}[t]{0.19\linewidth}
    \centering
    \includegraphics[width=\textwidth,trim=0.75cm 0.75cm 0.75cm 2.5cm,clip]{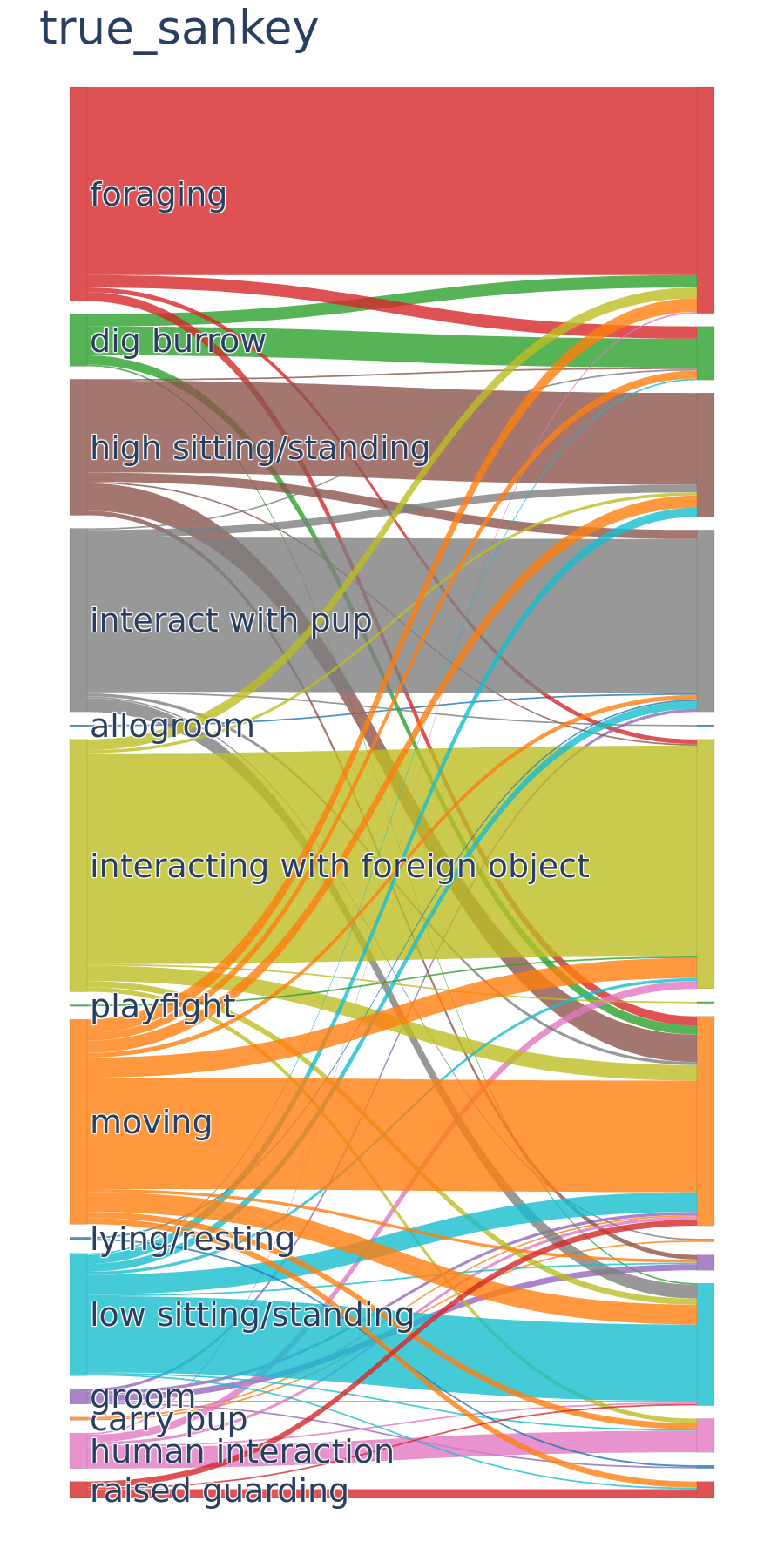}
    \caption{Ground truth.}
    \label{fig:ground_truth_sankey}
  \end{subfigure}
  \hfill
  \begin{subfigure}[t]{0.19\linewidth}
    \centering
    \includegraphics[width=\textwidth,trim=0.75cm 0.75cm 0.75cm 2.5cm,clip]{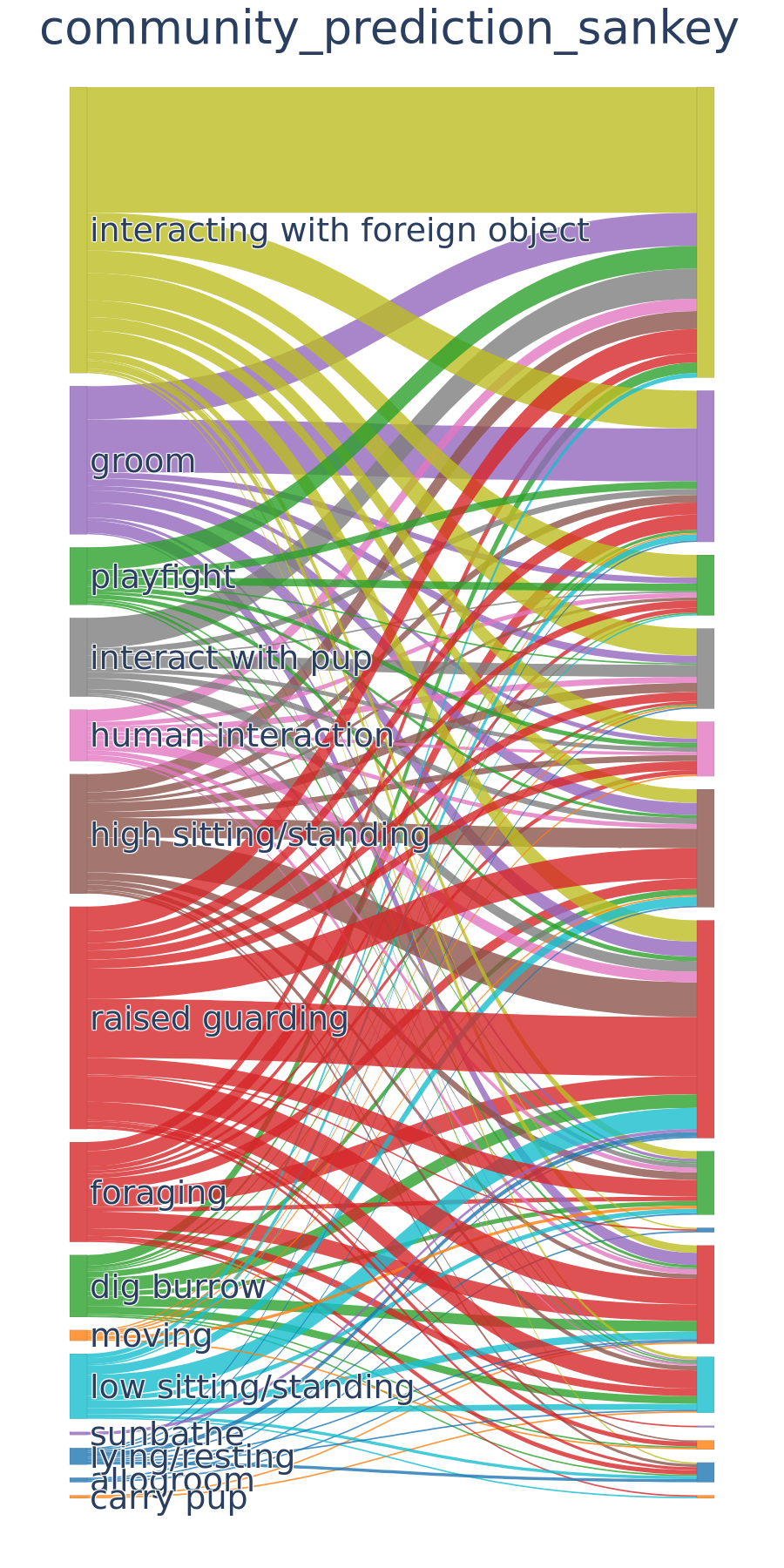}
    \caption{PCMCI$_C$.}
    \label{fig:pcmci_sankey}
  \end{subfigure}
  \hfill
  \begin{subfigure}[t]{0.19\linewidth}
    \centering
    \includegraphics[width=\textwidth,trim=0.75cm 0.75cm 0.75cm 2.5cm,clip]{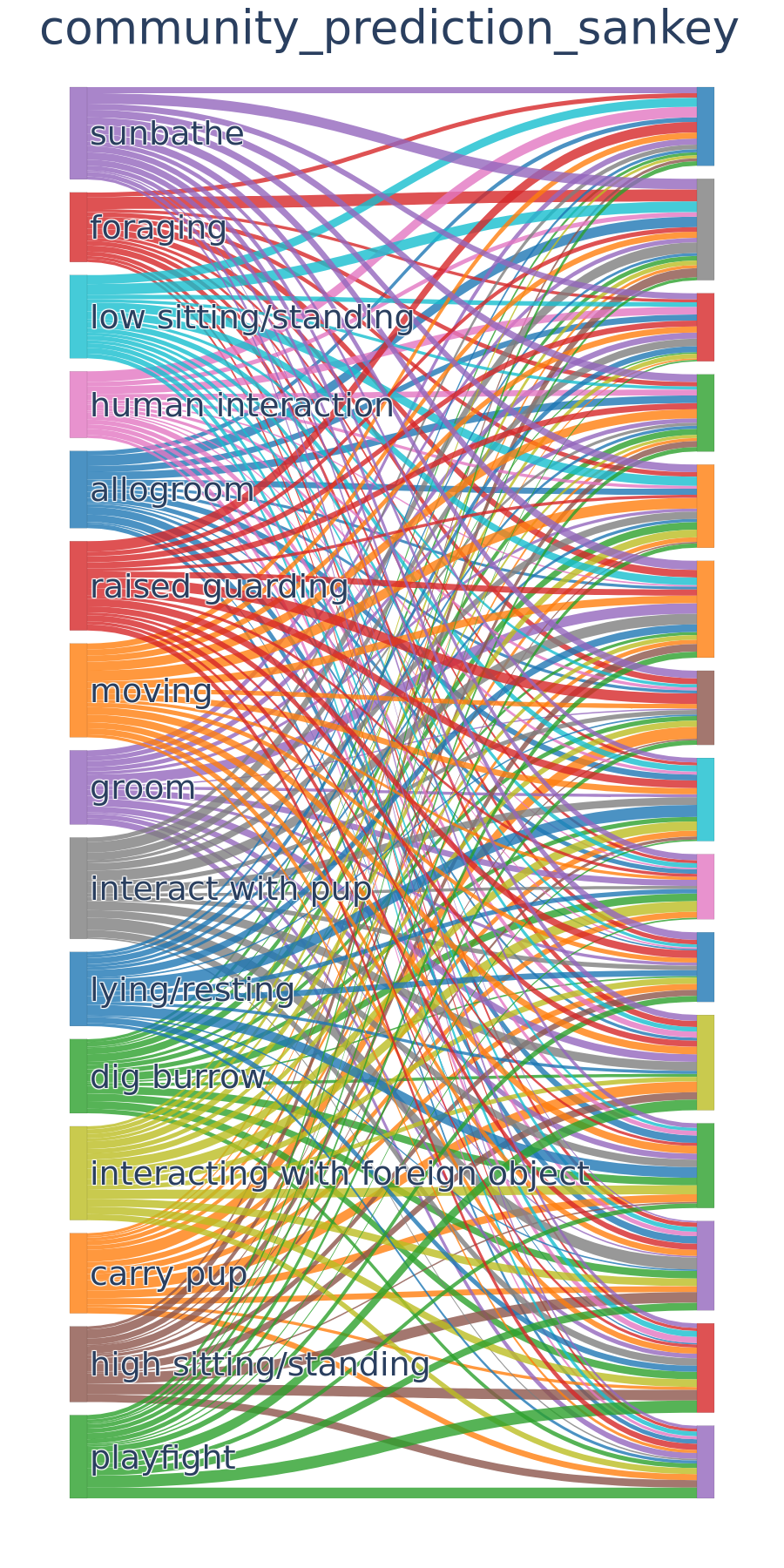}
    \caption{PCMCI+GCN$_C$.}
    \label{fig:pcmci_gcn_sankey}
  \end{subfigure}
  \hfill
  \begin{subfigure}[t]{0.19\linewidth}
    \centering
    \includegraphics[width=\textwidth,trim=0.75cm 0.75cm 0.75cm 2.5cm,clip]{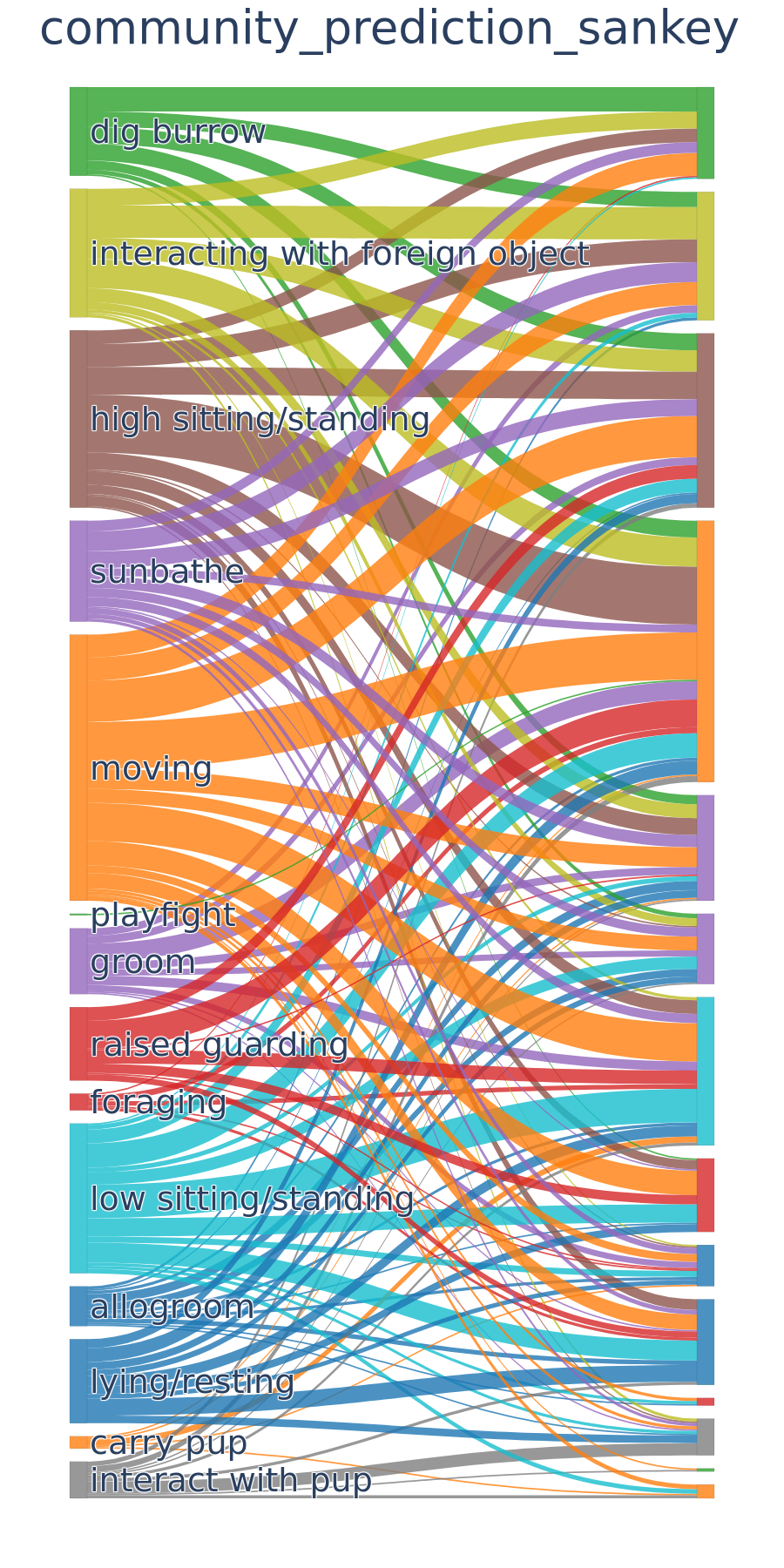}
    \caption{LSTM$_C$.}
    \label{fig:lstm_sankey}
  \end{subfigure}
  \hfill
  \begin{subfigure}[t]{0.19\linewidth}
    \centering
    \includegraphics[width=\textwidth,trim=0.75cm 0.75cm 0.75cm 2.5cm,clip]{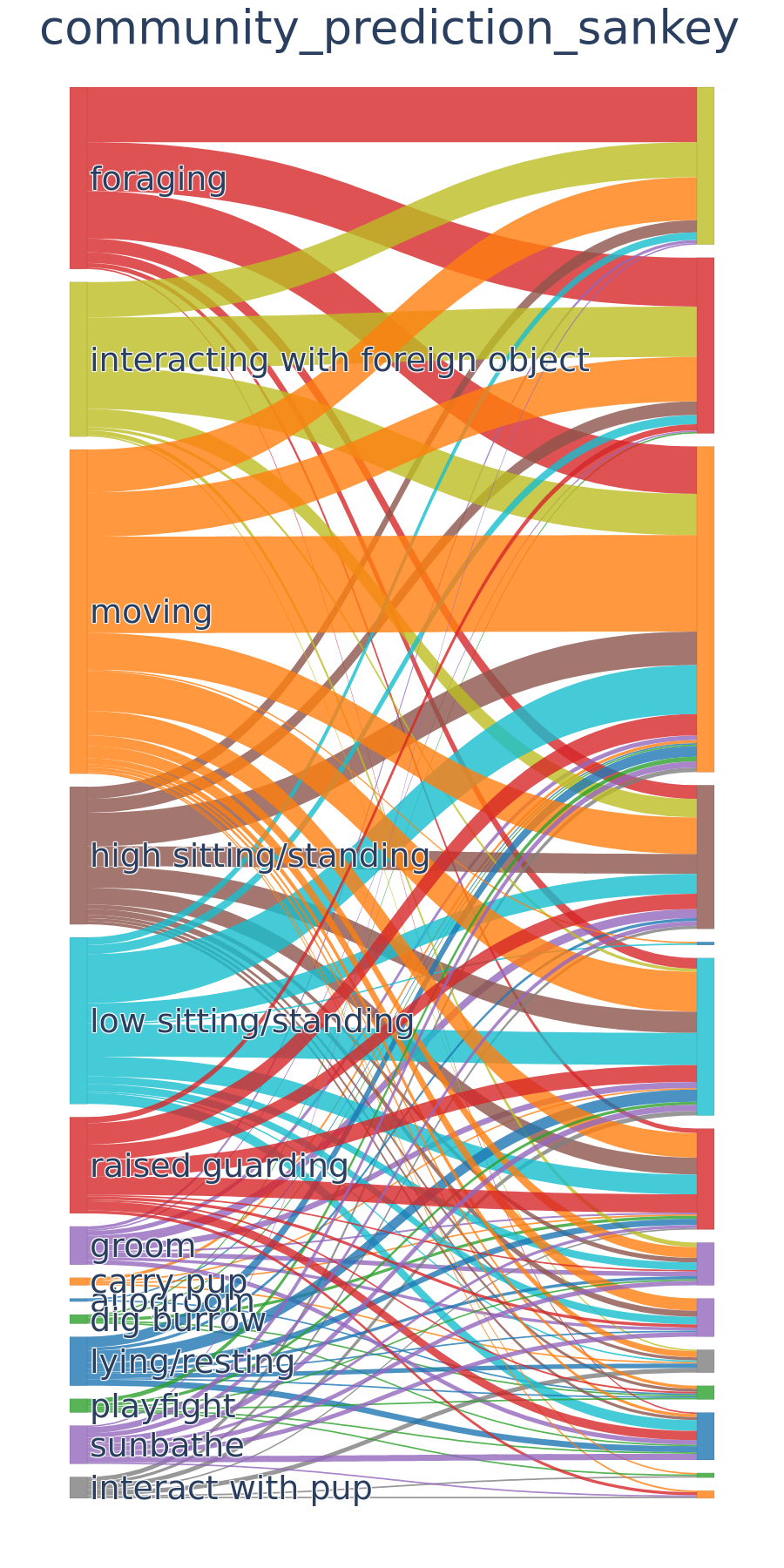}
    \caption{Transformer$_C$.}
    \label{fig:tf_sankey}
  \end{subfigure}
  \hfill
  \caption{Sankey diagram of the behaviour flows. The flow diagrams are shown for the series generated by a group of simulated meerkats. }
  \label{fig:sankey}
\end{figure*}

\subsection{Group-Level Simulation}

In this section, we study the prediction abilities of our model at the group-level. All individuals in the group are modelled such that the behaviour $s_t^{(k)}$ of one individual $k$ will affect the context information $C_{t+1}^{(j \neq k)}$ of the other individuals. We labelled the time-series generated for an entire group with a \textit{$_C$} to differentiate them from the time-series for an individual. 

\paragraph{Discrimination}

As for single individuals, we quantitatively evaluate the similarity between the true and the simulated data. Table \ref{tab:discriminator_results} shows the results. We focus on the second half of the table. The prediction models achieved better performance at the group-level when compared to single-individual time-series. This indicates that the behaviours do not collapse under the influence of the other predictions and maintain their initial proximity to the true behaviour distribution. Moreover, the influence of the modified context improves the quality of the generation. The interactions between individuals seem to be of paramount importance for accurately modelling their behaviours. As in the single-individual case, the LSTM discriminator distinguishes the true and counterfeit data better than the Transformer discriminator, with a lower standard deviation. PCMCI+GAT$_{\mathcal{G}C}$ and PCMCI+GCN$_{\mathcal{G}C}$ achieve the best results with the LSTM and Transformer discriminators, respectively. Overall, the GNN models achieve similar or better performances than the baselines, with the exception of PCMCI+GATv2$_{\mathcal{G}C}$ with the Transformer discriminator. PCMCI even achieves the second best score with this discriminator. This is consistent with the observations made in the individual-level case, although causal models take better advantage of group-level generation than the baseline models.

\paragraph{Visualisation}

As for single individuals, we visualise the clusters generated for short sequences of behaviours. The results are shown in Figure \ref{fig:clusters}. There are no apparent differences between the individual- and group-generated data. The samples generated at the group-level do not form organised structures and express patterns similar to their individual-level counterparts. In addition, we visualise the Sankey diagrams of the generated series in Figure~\ref{fig:sankey}. A Sankey diagram measures the flow between processes. In our case, it corresponds to the distribution of the behaviour transitions, i.e. what are the most likely behaviours one individual might take if in the current state. The ground truth Sankey diagram greatly differs from the other diagrams. The majority of the transitions are directed towards the same behaviour, and few transitions actually lead to other behaviours. This is consistent with the observations made in Section \ref{sec:dataset}. In contrast, the simulated series are more volatile. One behaviour often leads to new behaviours. LSTM$_C$ and Transformer$_C$ exhibit similar patterns. In particular, the  "moving" and "interacting with foreign object" behaviours occur more often than in the ground truth data.
One potential explanation is overfitting to the most observed behaviours in the training data to maximise the prediction accuracy. As shown in Figure \ref{fig:distrib-data}, "moving" and "interacting with foreign object" are frequently observed behaviours. Interestingly, the "foraging" behaviour has little input and output flows, despite being often observed in the ground truth. PCMCI$_C$ shows patterns similar to those of the baselines, although the behaviours with the highest flows differ. For instance, "raised guarding" has a high input flow, despite being scarcely observed in the data. This is not surprising, as the model aims to maximise the strength of the causal dependencies and not the likelihood of the behaviour transitions. However, there seems to be an important overlap between these two objectives. 
Finally, PCMCI+GCN$_C$ shows a very different pattern from the other diagrams. The behaviours are balanced and the model does not provide higher flows to some behaviours. This can be a possible explanation for the highest discrimination scores despite achieving a prediction performance similar to that of the baselines.

\subsection{Explainability}
\label{sec:explainability}

\begin{figure}[t]
    \centering
    \includegraphics[width=0.9\linewidth,trim=4cm 12cm 4cm 12cm,clip]{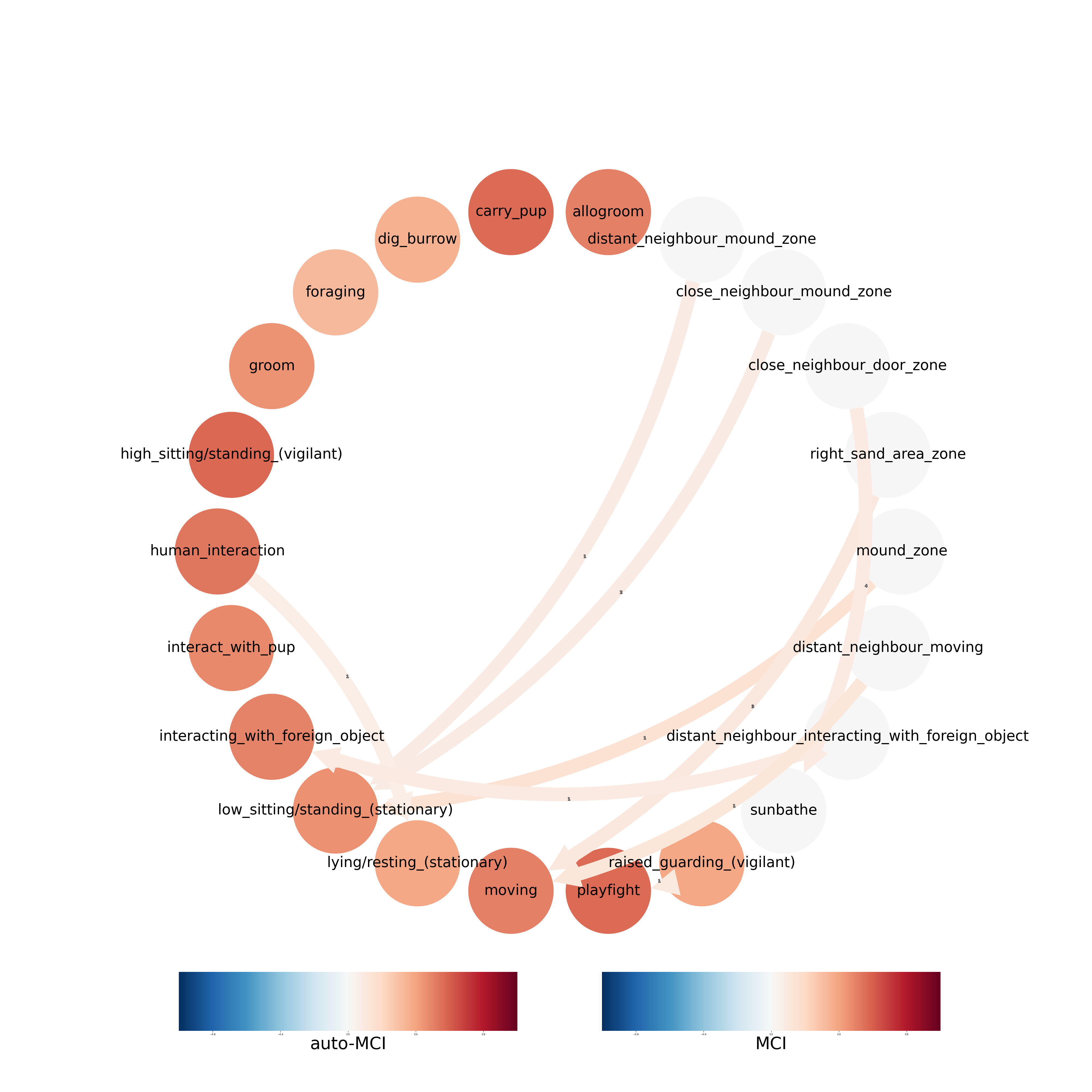}
    \caption{Causal graph generated by PCMCI. Only edges directed to state variables and with the highest causation strength are shown. Context variables with no directed edges with high strength to state variables are hidden.  }
    \label{fig:causal-graph}
\end{figure}

The PCMCI algorithm generates a causal graph that links cause variables to the variables affected by them. Such causal graph allows the building of a sparse agent model with greater interpretability than fully-connected models. It can provide insight into what behaviours can lead to another, which outcomes are not possible, and what context causes can influence one's behaviour. Figure \ref{fig:causal-graph} shows an example of causal graph generated by PCMCI. Only the links towards the state variables and with the highest causation strength are displayed. For instance, a "raised guarding" agent might start to "playfight" in the following timesteps; having neighbours "moving" or "interacting with a foreign object" can lead an agent to follow them. Further in-depth investigation into the structures of the causal graphs may lead to discoveries on the topic of interest, but this is beyond the scope of this paper.

\section{Conclusion}

Understanding the complex behaviours of social animals from observations is an important task encompassing numerous fields of study and applications, ranging from the welfare of animals in zoos to the understanding of animal social structures. This is a challenging task as the data collection process is expensive and may not cover the full scope of the causes affecting an individual's behaviour, in particular, the mental states of an individual are unknown.

In our work, we tackle the problem of modelling the behaviours of a group of meerkats interacting together in a zoo environment. We use causality theory and graph neural networks to build an interpretable prediction and generation engine. We show that our method can compete with and outperform standard deep learning models with a higher number of parameters. In particular, the proposed method can generate more realistic simulation data than the baselines. This work is a first step in the modelling of social animals, and many challenges are yet to be explored. This paper uncovers some limitations of the proposed and current models, such as the lack of information (in Shannon's definition) learned by the models and the large difference between the ground truth and simulated samples. We highlight a discrepancy between accuracy performance at the statistical level, and accurate modelling of the inner mechanisms of the agents.

In our future work, we will focus on learning new behaviours instead of relying on a pre-defined set. We will combine our current method with dynamical models to learn the movements of the agents in addition to their behaviours and study how it affects the level of information learned by the models. Finally, we will use Inverse Reinforcement Learning to model the long-term policies and rewards that drive the agent's behaviours. 
~\\


\balance
\bibliographystyle{ACM-Reference-Format} 
\bibliography{references/introduction, references/causal_modelling, references/application, references/related_work}


\end{document}